\documentclass[AER,floats]{AEA}
\usepackage{natbib}
\setcitestyle{authoryear,square}
\draftSpacing{1.5}
\usepackage{amssymb}
\usepackage{nicefrac}
\usepackage{gensymb}
\usepackage{enumerate}
\usepackage{amsmath}
\usepackage{mathptmx}
\usepackage[english]{babel}
\usepackage{graphicx}
\usepackage{floatrow}
\usepackage{wrapfig}
\usepackage{afterpage}
\usepackage{color}
\usepackage{caption}
\usepackage{subcaption}
\usepackage{setspace,multirow}

\newtheorem{definition}{Definition} 

\begin{document}

\title{Lattice Studies of  Gerrymandering Strategies}

\author{Kyle Gatesman and James Unwin\thanks{Gatesman: 
Thomas Jefferson High School for Science and Technology, Braddock Rd, Alexandria, VA 22312, USA. 
Unwin: University of Illinois at Chicago, Chicago, IL 60607, USA (email: unwin@uic.edu). We would like to thank T.~Khovanova for helpful interactions  and F.~I.~Schaposnik Massolo  and L.~P.~Schaposnik for comments on a draft of the manuscript. This research was undertaken as part of the MIT-PRIMES program. JEL Codes:  D72, H10, K00.}}
\date{ \today}



\begin{abstract}

We propose three novel gerrymandering algorithms which incorporate the spatial distribution of voters with the aim of constructing gerrymandered, equal-population, connected districts. Moreover, we develop lattice models of voter distributions, based on analogies to electrostatic potentials, in order to compare different gerrymandering strategies.  Due to the probabilistic population fluctuations inherent to our voter models, Monte Carlo methods can be applied to the districts constructed via our gerrymandering algorithms. Through Monte Carlo studies we quantify the effectiveness of each of our gerrymandering algorithms and we also argue that  gerrymandering strategies which do not include spatial data lead to (legally prohibited) highly disconnected districts. Of the three algorithms we propose, two are based on different strategies for packing opposition voters, and the third is a new approach to algorithmic gerrymandering based on genetic algorithms, which automatically guarantees that all districts are connected. Furthermore, we use our lattice voter model to examine the effectiveness of isoperimetric quotient tests and our results provide further quantitative support for implementing compactness tests in real-world political redistricting. 

\end{abstract}

	\maketitle
	

	Representative democracies must necessarily group constituents into voting districts  by partitioning larger geographical territories. Gerrymandering is the act of purposely constructing voting districts which favour a particular electoral outcome.  In the United States the power to draw district lines within a state belongs to the state legislature or districting commission. Thus, self-interested politicians with this authority could gerrymander -- manipulate the district lines of their territory -- to maximize the electoral outcome for their own party. Gerrymandering for political gain is morally questionable as it reduces the power of the electorate, and this practice is not restricted to any political party or country.  Indeed, the Supreme Court of the Untied States has recently heard two gerrymandering cases, the first  \cite{WvG} concerned the 2011 redistricting plan for Wisconsin due to Republican legislators, and the second \cite{BvL} was regarding changes made to the boundaries of Maryland's 6${}^{th}$ district by the Democratic Party.  Furthermore, in principle, there are instances in which elaborate redistricting could be applied with benevolent intent, such as  ensuring the proper representation of minority groups (based on ethnicity, religion, or other identifiers) which are not spatially localized. Such majority-minority districts have also been the focus of Supreme Court hearings, e.g.~\cite{SvR} and  \cite{MvJ}.

\newpage
	
Political gerrymanderers aim to maximize the number of districts in which constituents of an opposing party will assuredly lose the majority vote, thereby minimizing the opponent's political influence.  
However, districts are commonly required to conform to certain general requirements:
	\begin{itemize}
		\item Connectedness: Each district must comprise a single, connected region.

		\item Uniformity: All districts in a territory must have approximately equal populations.

		\item Shape: Districts should be generally compact, but legal stipulation is limited.
	\end{itemize}

\begin{wrapfigure}{R}{0.55\textwidth}
  \ffigbox[\FBwidth]
  {\vspace{-3mm}
    \caption{Sample 5$\times$5 territory with two different district allocations.   \label{Fig1}} }
  {    \includegraphics[width=0.85\textwidth]{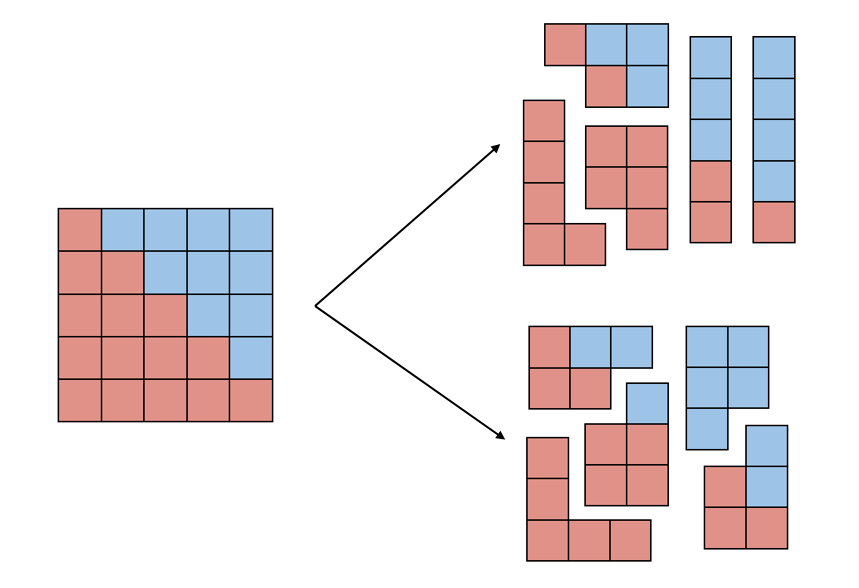}  }
\end{wrapfigure}
	
Despite these requirements, clever redistricting can have significant consequences. Consider an election involving two parties, which we label {\em red} and  {\em blue}, and a territory which can be modelled as a 5 $\times$ 5 grid. Each of the 25 unit squares denotes a territorial unit, and its colour represents the overall party affiliation of its voters. For the purpose of this simple example, we will assume a uniform population (thus each unit has equal voting weight) and a voter preference such that 60\% (40\%) of the units favour red (blue). 

Given the split in voter preference, an impartial districting into five districts should be expected to yield three red-majority districts and two blue-majority districts. However, as illustrated in Figure \ref{Fig1}, it is possible for the blue party to gerrymander the territory so that it wins three out of the five districts, thereby winning a majority of districts. Conversely, the red party can construct four red-majority districts instead of three. Thus, an entity with the power to set the district lines can potentially arrange for whatever result it desires if unconstrained by other considerations. This illustrates, a simple, but powerful, gerrymandering strategy in which  opposition voters are  ``packed'' into districts in a manner which wastes the voting power of opposition supporters. 
	
		The main pursuit of this work is to construct algorithms which take a  distribution of voters on a lattice and returns  a set number of gerrymandered, equal-population, connected (or mostly connected) districts. Lattice studies of redistricting can clearly provide a great deal of insight, and thus we use our model to quantify some general statements concerning gerrymandering. In particular we use our lattice population models to compare gerrymandered districts to geometrically constructed `fair' districts and examine how this changes the net vote in each district and the overall election result, in order to quantify to what extent gerrymandering is advantageous to the proponent party.  Moreover, by applying common measures of gerrymandering to the districts generated via our algorithm, we are able to provide a quantitative assessment of whether these measures can detect and potentially constrain gerrymandering.

	An influential paper of  \cite{FH} systematically explored algorithmic approaches of ``packing'' and ``cracking'' voters into districts, arriving at the mantra ``sometimes pack, but never crack'', and propose a novel packing procedure for strategically gerrymandering a territory. Despite providing a number of excellent insights, the gerrymandering algorithm proposed in \cite{FH} entirely neglects the spatial distribution of voters and thus generally leads to highly disconnected voting districts. In the present paper we develop a lattice model which encodes population distributions and voter preferences. Using this lattice model we study the spatial profile of the aggressive gerrymandering strategy outlined in \cite{FH} and shall show that it generally leads to highly disconnected districts.

Specifically, in this work we study four strategies for gerrymandering. The first strategy is an implementation of the \cite{FH} method which references a lattice voter distribution. The latter three strategies are novel algorithmic approaches we propose here:

\begin{itemize}

\vspace{1mm}

\item {{\bf Friedman-Holden (FH) packing} (Section \ref{sec3b}):} Districts are formed from the most partisan voters from both parties, with a bias such that most districts favour the gerrymander's party. The algorithm does not required the districts to be connected.

\vspace{1mm}

\item {{\bf Spatially Restricted Friedman-Holden (SRFH) packing} (Section \ref{srfh}):} The FH packing strategy is adapted to ensure almost all districts are connected. 

\vspace{1mm}

\item {{\bf Saturation packing} (Section \ref{sat}):} Opposition voters are packed into a small number of districts, skewing the partisan bias in the majority of districts.

\vspace{1mm}

\item {{\bf Genetic gerrymandering} (Section \ref{sec4b}):} Starting from sets of random districts, we iteratively mutate these district configurations, in order to maximize some predefined fitness function. Choosing the fitness function appropriately can yield both fair or gerrymandered sets of districts. 

\end{itemize}

To some extent the algorithms developed here are driven by two competing goals
\begin{itemize} 
\item[{\em i)}] Maximising the number of districts won by the gerrymanders party; 
\item[{\em ii)}] Aiming for connected (or mostly connected) districts. 
\end{itemize} 
Indeed, it is a common legal requirement that voting districts are a single connected region, however, as we show in Section \ref{sec3b}, the  approach of \cite{FH} leads to all districts being highly disconnected. In contradistinction, in the genetic gerrymandering algorithm we develop here all districts are guaranteed to be connected, and in the Saturation and SRFH packing strategies only the final district remains disconnected. In the latter case, the final district typically has only a small number of distinct pieces and connectivity can often be achieved through minor swaps between districts. Hence, providing a significant improvement on the algorithmic gerrymandering strategy of \cite{FH}. 

	We note here that there is a sizeable body of existing literature focusing on minimizing, optimizing, and detecting gerrymandering. In particular, several groups have proposed methods to construct fair districts, which are population equal and non-partisan  (for instance~\citep{Sherstyuk,Bard}), or maximally gerrymandered, which favour a particular outcome (see e.g.~\citep{Sherstyuk,FH,PT2,Apollonioa}.  Additionally, several studies have presented a range of geometric tests or measures, such as voting district \textit{compactness} and \textit{convexity}, to detect gerrymandering, for example \citep{Roeck,Schwartzberg,Oxtoby,Young,Niemi,Polsby,Chambers,Hodge,Wang,Giansiracusa,Duchin,Warrington}.

	 This work is structured as follows: In Section \ref{sec3}, we outline a new procedure for generating lattice models of population and voter distributions. In Section \ref{sec3b} we outline a specific model of aggressive gerrymandering, proposed by \cite{FH}, and using our lattice models we demonstrate that this leads to disjointed districts. Subsequently, in Section \ref{sec4}, we outline two packing algorithms, one of which is based on similar principles to Friedman and Holden's approach, and both of which take into account spatial information regarding voters. In Section \ref{sec4b} we present a further algorithmic gerrymandering strategy based on genetic algorithms, with the distinct advantage that it automatically outputs connected districts. In Section \ref{sec5} we apply our codes to generate a number of gerrymandered territories, presenting both instructive examples and Monte Carlo studies which quantify the impacts of gerrymandering. Finally, in Section \ref{sec6}, we give a summary of results, a discussion of their implications, and suggest potential directions for subsequent studies.  Our Python codes which implement the algorithms discussed herein are provided online.


\section{Modelling Voter Distributions on Lattices}
 \label{sec3}

A manner of generating large sets of quasi-random population and voter distributions can provide a flexible tool for studying the general features of population subdivisions and gerrymandering. Abstracting away from purely data-driven studies of voter distributions can allow both more general analyses and more specialized studies depending on how one implements the model. In this section we outline an elegant manner of constructing models of a voter distribution. Specifically, we propose to study a population distribution which is modelled on a binomial distribution (generated by a walker algorithm), which approximates well a discretized Gaussian distribution with random fluctuations. Such a quasi-Gaussian distribution is a good model for a city or town in the absence of natural boundaries (which breaks the spherical symmetry). We then superimpose a spread of partisan bias on this population. 
Whilst the notion of modelling voters via lattice distributions has been previously explored in applications of statistical physics to sociopolitical research e.g.~\citep{Chou,Wall,Castellano}, to our knowledge there are no previous studies which apply lattice techniques to assess the viability of specific gerrymandering strategies  for a given voter distribution.
 
 		\subsection{Lattice Models of Population Distributions}
	\label{2.1}

In this sections we shall define the key concepts that will be used throughout the paper, which concern modelling geographical regions with a population of voters, which we refer to as territories. In most representative democracies, it is common to split territories into small indivisible cells called territorial units (such as census units), with each unit containing a potion of voters. In this work we model territories as lattices:
\begin{definition}
 A {\bf territory} $S$ is a square lattice in $\mathbb{Z}^2$, where each lattice site $(i,j)$ defines  a {\bf  territorial unit} $T_{i,j}$ carrying a population value $P_{i,j}\in\mathbb{N}$ and a voter preference $v_{i,j}\in(-1,1)$. The total population of the territory is defined as $P_{S}=\sum_{i,j} P_{i,j}$.
 \end{definition} 
 
 We shall call a  population distribution on a territory $S$ a set of fixed values for all $P_{i,j}$ and call a ``voter'' (or ``partisan'') distribution a set of fixed values for all $v_{i,j}$. We call $S$ equipped with a population distribution a  ``population model'' and $S$ equipped with both a population and voter distribution will be referred to as a  ``voter model''.
\begin{definition} 
Given a territory $S$, a  {\bf district} $D$ is a finite union of territorial units, i.e.,  $D=\cup_{(i,j)\in I} T_{i,j}$ for an index set $I$. 
The district population is defined as $P_D=\sum_{(i,j)\in I} P_{i,j}$ and district voter preference is $N_D=\sum_{(i,j)\in I} v_{i,j}$. 
 \end{definition}
In contrast to arbitrary graphs, lattice territories can be efficiently manipulated and  are ideal for our analysis, since the distributions input to our algorithms and the districts output can all be represented as square matrices.
Furthermore, the lattice structure provides intuitive notions of adjacency and connectedness between the territorial units:
\begin{definition} 
Territorial units at $T_{i,j}$ and $T_{k,l}$ are said to be {\bf adjacent} if $i=k\pm1$ and $j=l$, xor (exclusive or) $j=l\pm1$ and $i=k$. 
\label{def77}\end{definition} 
\begin{definition} 
A territorial unit $T_{i,j}$ is {\bf reachable} from $T_{k,l}$ if there exists a sequence of adjacent territorial units beginning at $T_{i,j}$ and ending at $T_{k,l}$.
 \end{definition} 
Given the above definitions of adjacent units and reachable units, we can express a simple notation of district connectedness:
\begin{definition} 
A district $D$ is {\bf connected} if any $T_{i,j}\in D$ is reachable for every $T_{k,l}\in D$.
 \end{definition}

Since we are interested in cases where the territory is partitioned into a set of equal population districts,  we introduce the following definition:
\begin{definition} 
A  {\bf valid districting} is a set of $n$ disjoint districts $\{D_i\}$ for $1\leq i\leq n$ such that $S=\cup_{i\leq n} D_i$ and for fixed $t\in\mathbb{R}$ one has $-t\leq|D_i|-|D_j|\leq t$ for $1\leq i,j\leq n$. 
\label{def7}
 \end{definition}
The quantity $n$ denotes the total number of districts in $S$. We call $t$ the population threshold, which allows for small variations in population between districts, whilst requiring approximately equal district populations. Throughout this work we will take $t \sim 0.01\times P_S/n$, such that differences between districts are percent-level.

\newpage

	Since it is of interest to consider population distributions which well model real world situations, here we shall focus on a quasi-Gaussian distribution of population much as would be appropriate in a large city in which the population is highly dense towards the centre and becomes diffuse at large radial distances. To approximate a Gaussian population spread  with random fluctuations, we implement a {\em walker function} (see e.g.~\cite{Shiffman}) on a $m\times m$ lattice with $m\in\mathbb{Z}^{\rm odd}$ with the central lattice site designated $(0, 0)$. The walker function is essentially a simple agent-based model (see e.g.~\cite{Macal}) which undergoes time step evolution. In this case an agent is an object associated to a single lattice site at a given time step and the walker function is a set of probabilistic rules which determine how the spatial location of agents evolve between time steps. The agent represents an individual of the population and thus the probabilistic evolution of agents leads to random fluctuations in the population distribution. We will exploit these random fluctuations in the population distribution to implement Monte Carlo methods later in this work.
 	
\begin{wrapfigure}{R}{0.4\textwidth}
\vspace{10mm} 
  \ffigbox[0.85\FBwidth]
  {    \caption{Example population distribution generated by the walker algorithm. Colour intensity indicates population density.   \label{FigPop}} }
  {    \includegraphics[width=0.75\textwidth]{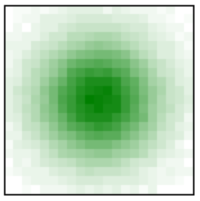}  }
\end{wrapfigure}
	
For a given territory $S$ one can construct a population model with total population $P_S$ via the walker function, as detailed below. We take a $m\times m$ lattice and consider $P_S$ agents with the following starting distribution (at time step $t=0$)
 \begin{equation}
P_{i,j}\big|_{t=0}=
\left\lbrace
\begin{array}{ll}
P_S
 &~~ \text{for}~(i,j)=(0,0) \\[10pt]
0 &~~ \text{otherwise}
\end{array}
\right.~.
 \end{equation}
Thus, each integer unit of population is associated to an agent on the lattice and, prior to evolution via the walker function, the whole population of the territory is located at the central lattice site. 

During each time step an agent can move with fixed probability to any lattice site adjacent to its current location with equal probability, with the restriction that agents remain within the $m\times m$ lattice. Each agent is allocated a fixed number of moves, and move counts across all agents follow a normal distribution centred at the minimum number of moves needed to reach $(m,m)$ from $(0,0)$. Once an agent has taken its prescribed number of moves, it remains in its terminal unit. 

After all agents have taken their prescribed moves the walker function outputs the number of agents at each lattice site $(i,j)$ and this is identified with the population $P_{i,j}$ of the territorial unit $T_{i,j}$. The walker function provides a value for $P_{i,j}$ for each $T_{i,j}\in S$ and thus defines a population model (but not a voter model, since the $v_{i,j}$ remain undetermined at this stage). The population spread due to this algorithm well approximates a two-dimensional Gaussian distribution; an example is shown in Figure \ref{FigPop}.

\subsection{Modelling Elections on the Lattice}
	\label{2.2}

	With a working lattice model of population distributions we next introduce a flexible manner of introducing a voting distribution within the population which shall lead to the premise of voting and electoral events. 
	 \begin{definition}
The \textit{\bf proponent} (\textit{\bf opponent}) is the party which benefits (loses) from gerrymandering. A territorial unit $T_{i,j}$ is a \textit{proponent unit} if $v_{i,j} > 0$, an \textit{opponent unit} if $v_{i,j} < 0$, or \textit{neutral} if $v_{i,j} = 0$.
	\end{definition} 
Without loss of generality, we let the proponent party corresponds to positive extremity, i.e.~$v_{i,j}>0$, and designate this the ``red'' party. The opponent party we designate the ``blue'' party.
We assume all voters cast ballots, so $v_{i,j}>0$ corresponds to the average vote in territorial unit $T_{i,j}$  favouring the proponent. We also assume  that the gerrymander knows the values $v_{i,j}$ with certainty,  although this could be relaxed. 
\begin{definition}
The {\bf net territory vote} (or popular vote) of a territory $S$ is the sum $N_S:=\sum_{i,j}v_{i,j}$.  
A territory is said to be {\bf balanced} if $N_S\approx0$.  For a district $D=\cup_{(i,j)\in I} T_{i,j}$ in $S$, the {\bf district vote} is $N_{D}=\sum_{(i,j)\in I} v_{i,j}$.
\end{definition}
  We say that the red party  wins the popular vote in $S$ if $N_S>0$; conversely, blue wins (red loses) the \textit{popular vote} if $N_S<0$. However, rather than the popular vote of the whole territory, what is typically most important is the district-wise overall vote. 
 Similar to the popular vote, for $n$ districts, we say that red wins the district vote of district $D_k$ for $1\leq k\leq n$ if $N_{D_k}>0$ and we say blue wins (red loses) the district vote if $N_{D_k}<0$.

The distinction between the district vote and the popular vote implies that a given party can lose the latter, while securing the most districts. Typically the most important outcome is the number of districts won by each party. Thus, gerrymandering fundamentally exploits the differences between the local and global properties of distributions.
Realistic numbers of districts $n$ in a given territory range from 2 to $\mathcal{O}(10)$; for example Hawaii has only 2 congressional districts, while California has 53. In our later example and statistical studies we will typically take $n=5$. 

Since elections are dynamic and inherently uncertain, the gerrymanderer risks losing if the plan is to win a district by only a single vote. Introducing a vote threshold $w$ ensures that the gerrymanderer's party wins a given district with a minimum margin.\footnote{One could define $w_k$ district-by-district, such that the requirement is weaker or stronger depending on the district.} We call a district ``safe'' for red if  $N_{D_k}>w$, for a fixed prescribed $w\in\mathbb{Z}$ which is called the vote threshold (conversely,  $N_{D_k}<-w$ is safe for blue).  We typically take $w\sim 0.01\times P_S/n$, for $n$ districts, such that safe districts favour the proponent with a margin of at least 1\%.

\subsection{Modelling Voter Preference Distribution}

	In what follows we shall introduce a novel procedure for generating smooth distributions of voter preferences. We implement voter preference in terms of a number of specified points of peak partisan bias, `sources', with voter preference falling towards neutrality away from these peaks. Using an analogy to the theory of electrostatic potentials and inverse power laws, we define sources of partisan bias via the properties of point charges, as follows.
		
\begin{wrapfigure}{R}{0.3\textwidth}
  \ffigbox[0.9\FBwidth]
  {  \vspace{-7mm}
\caption{Example voter distribution for two source points.   \label{Fig4}} }
  {    \includegraphics[width=0.8\textwidth]{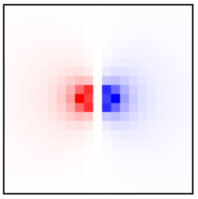}  }
\end{wrapfigure}

\begin{definition}
	Given a territory $S$, a {\bf source point} is a pair $E_{i,j}=\{(i,j),e\}$ characterized by its location $(i,j)\in S$ and the magnitude $e\in\mathbb{R}$ of the source. We require that any set of source points gives $|v_{i,j}|\leq1$ $\forall T_{i,j}\in S$.
	\end{definition}
To match with our previous (arbitrary) assignment that $v_{i,j}>0$ corresponds to the red party, we call $E_{i,j}$ a red source when $e>0$  and a blue source when $e<0$. 
Source points can be located at arbitrary lattice sites, and the voter preference at a given territorial unit $T_{i,j}$ is a function of the distance $d$ from these source points, following a $1/d$ power law:

	\begin{definition}
	The {\bf vote contribution} $\Delta_{k,l}$ from a source by $E=\{(i,j),e\}$ to the net vote of territorial unit $T_{k,l}$, where $(i,j)$ and $(k,l)$ are separated by distance $d(T_{i,j},T_{k,l})$ is 
	\begin{equation}
	\Delta_{k,l} =  \frac{e}{\max[1, d(T_{i,j},T_{k,l})]}~,
	\end{equation}
\end{definition}
where the distance function (the metric) is taken to be
	\begin{equation}\label{eqd}
d(T_{i,j},T_{k,l}) := \sqrt{(k - i)^2 + (l - j)^2}~. 
	\end{equation}
Given $\alpha$ source points $\{E_\alpha\}$ for $1\leq \alpha \leq m$, whose positions are chosen independently, we denote their contribution to $v_{i,j}$ as  $\Delta_{i,j}^{(\alpha)}$  and 
	\begin{equation}
	\label{vij}
	v_{i,j} = \sum_m\Delta_{i,j}^{(m)}~.
	\end{equation}
From the above definition the vote contribution from a given source falls linearly with distance $d$ from the source.\footnote{In principle one could study other power laws or consider sources each with different $d$ dependencies.} Because of the $1/d$ power law, the source points typically represent local maxima of the voter preference, with sign$(e)$ indicating the favoured party. In this model, a balanced territory requires at least one blue and one red source, so we are generally interested in scenarios with two or more sources.

	Additionally, we note that in principle two source points  could be located at the same lattice site, 
	\begin{equation}
E_1=\{(i,j),e_1\}
 \quad {\rm and}  \quad  E_2=\{(i,j),e_2\}~,
	\end{equation}
 in this case the two sources can always be replaced with a single source: 
	\begin{equation}
	\{E_1,E_2\}\leftrightarrow E_{1+2}=\{(i,j),(e_1+e_2)\}~.
	\end{equation}

\newpage
\subsection{Benchmark Models}

To summarize, given a territory $S$ we use the walker function of Section \ref{2.1} to fix the $P_{i,j}$  values  of $S$ and by designating a set of source points and referring to eq.~(\ref{vij}) we fix $v_{i,j}$ values of $S$, thus defining a voter model.
As an example, we assign lattice sites immediately left and right of the origin $(0,0)$ to serve as blue and red source points $\{E_{B},~E_{R}\}$ with  
		\begin{equation}
E_R=\{(-1, 0),1\} \quad {\rm and} \quad 	E_B=\{(1, 0),-1\}~.
\end{equation}
 The combination of the two dimensional quasi-Gaussian population distribution (as in Figure \ref{FigPop}) and the sources $\{E_{B},~E_{R}\}$ produces the voter distribution shown in Figure \ref{Fig4}. The colour intensity indicates the magnitude of the net voter preference $v_{i,j}$ and the centre of these coloured regions corresponds to the positions of the two source points.

In subsequent examples and statistical analyses presented throughout the remainder of this work we shall consider a number of specific benchmark voter models with a quasi-Gaussian population distribution and particular source point distributions, as given in Table \ref{tab0} below:

		\begin{figure}[b!]
		\includegraphics[width=0.75\textwidth]{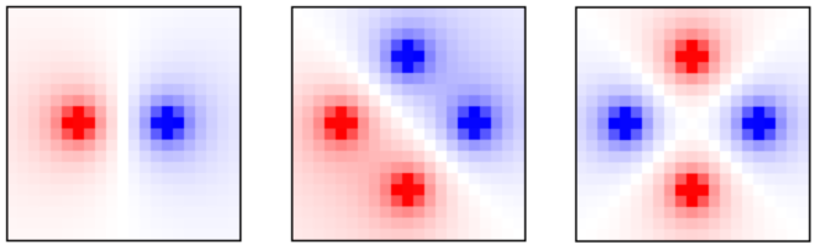}\\
		{Model \#2 \hspace{2cm} Model \#3 \hspace{2cm} Model \#4}
		\centering
		\caption{Visualization of voter distributions for models \#2 - \#4 of Table \ref{tab0}. \label{Fig4b}}
	\end{figure}

 \vspace{-1mm}
\begin{table}[H]
\begin{center}
\def\str{\vrule height10pt width0pt depth7pt}
\begin{tabular}{| c | c | c | c | c | c  | c | c | c | c | c | c | c | c | c  | c | c |  c | c | c | }
    \hline\str
Model \#  & $E_B$             & $E_B'$ & $E_R$             & $E_R'$  
   \\  
\hline
1 &  $((1,0),-1)$    & -          &  $((-1,0),1)$          &-    \\    
2 &   $((4,0),-1)$    & -          &   $((-4,0),1)$   &  -        \\
3 &  $((6,0),-1)$    &  $((0,6),-1)$  &  $((-6,0),1)$    &  $((0,-6),1)$      \\
4 & $((6,0),-1)$    &  $((-6,0),-1)$  &  $((0,6),1)$    &  $((0,-6),1)$            \\
\hline
\end{tabular}
 \vspace{-1mm}
\caption{Models with 2 or 4 source points for territories based on $21\times21$ square lattices. A dash indicates that the source point is  not included in a given model. \label{tab0}}
\end{center}
\end{table}
 \vspace{-3mm}

\noindent	
All of the benchmark models have balanced territorial votes: $N_S\approx0$. The voter distribution of model \#1 is illustrated in  Figure  \ref{Fig4} and models \#2- \#4 are shown in Figure  \ref{Fig4b}. These examples show that the method above can implement a variety of voter distributions.


\section{The Friedman-Holden Packing Strategy}
\label{sec3b}

	There are two fundamental strategies in algorithmic gerrymandering: \textit{packing} and \textit{cracking}. First, a gerrymanderer can dilute the voting power of the opponent party either by \textit{packing} the most concentrated opponent-voting subpopulations into a small number of districts. Second, one can \textit{crack} the most concentrated opponent population into several districts so that the most concentrated or extreme voting base for the opponent party never gains a majority. A strategic application of voter packing underlies the approach of \cite{FH}.

As an the example consider a gerrymanderer that favours the red party and whose goal is that $N_{D_k}>0$ in the maximum number of districts $D_k$ for $1\leq k\leq n$ in a given territory.  \cite{FH} considered a pseudo-normal voter extremity distribution and generated districts by simply partitioning the bell curve of the population by extremity.  The first district is formed by joining the most extreme subpopulations, i.e., the bell curve tails, so that {\em i)} their combined population is approximately the average district population, and {\em ii)} the right tail is sufficiently larger than the left tail. The latter condition signifies that the extreme right-party voters are sufficient to override the extreme left-party vote in their district. 

 The above process will, in essence, ``waste'' the opponent's strongest voting population in a district it cannot likely win. This process is repeated on the subsequent districts and the final district is composed of the remaining population. Thus, by construction, the later districts are comprised of mostly moderate voters and, for balanced territories, are typically won by the opponent party. This approach has a number of merits but suffers from the lack of spatial considerations. The Friedman-Holden method equates to unrestricted ``cherry picking'': the gerrymanderer has the freedom to select scattered population chunks for placement in the same category, as we demonstrate shortly. Notably, if even a single district is disconnected, the districting plan is typically legally prohibited. 

A useful measure of failure, is the number of connected components of each district:
 \begin{definition} 
A {\bf connected component} $C\subseteq D$ of a district $D$ is a (non-empty) set of territorial units in $D$ such that given a territorial unit  $T_{i,j}\in C$, another territorial unit $T_{k,l}$ also lies in $C$ if and only if $T_{i,j}$ is reachable from $T_{k,l}$. 
 \end{definition} 
A district can be decomposed into its set of connected components $C_i$ and we shall write  $D=\cup_{i\leq r} C_i$, where $r$ is the number of connected components. If any territorial unit in $D$ is reachable from all other territorial units in $D$, then $r=1$ and we say that $D$ is connected. The number of connected components is important for the analysis of the spatial distributions arising through the algorithms studied. 

\subsection{Implementing Friedman and Holden's Algorithm on Lattice Territories}
\label{fhal}

\cite{FH} outline a  packing strategy which ignores the spatial data of the voter distribution. In order to demonstrate how this strategy leads to highly disconnected districts we shall reformulate the strategy of \cite{FH} for generating districts in terms of an algorithmic approach applied to a lattice voter model and we will refer to this algorithm as ``FH Packing''. Then by neglecting spatial data during the redistricting process, but tracking the positions of the territorial units allocated to each district, we can assess the connectivity of the districts constructed via FH packing.
	
	First, since voting districts are legally required to have comparable populations, we define a {\em target population} $P_{D\pm}$ to ensure that all districts have approximately equal populations. The target population is implemented using the population threshold $t$  (from Definition \ref{def7}), the total population $P_S$, and the number of districts $n$, as follows
\begin{equation}
P_{D\pm}:=\left(P_S/n\pm t\right)~.
\end{equation}
The value of $P_{D\pm}$ is computed before the algorithm is executed and each district should satisfy the following population condition
\begin{equation}
P_{D-}\leq P_{D_k}\leq P_{D+}~.
\label{pop}
\end{equation}
Also, the majority of district should satisfy the district win condition 
\begin{equation}
N_{D_k}>w~.
\label{dkw}
\end{equation}
Later districts, in particular the final district,  must have an opponent bias if the territory  is balanced. Only when the algorithm is satisfied with the composition of a given district will it proceeds to form the next district, until all $n$ districts are formed. 

To implement FH packing strategy  on a lattice territory  our algorithm iteratively assigns single territorial units to a district, one district at a time, such that the end result favours the proponent.  We call a territorial unit {\em unassigned unit} if it has not yet been assigned to a district and denote the set of unassigned units $U$. As  territorial units are assigned to districts by the algorithm, they are deleted from $U$.
 For a territory $S$ on a $m\times m$ lattice there are initially $m^2$ unassigned territorial units in $U$ and $n$ (empty) districts $D_k$ for $1\leq k \leq n$.  We implement the FH packing strategy  by first sorting the territorial units in order of decreasing net vote $v_{i,j}$, using a {\em quicksort} method \citep{quicksort}, and relabelling the elements of this ordered set $\{\hat{T}_1,\hat{T}_2,\cdots \hat{T}_{m^2}\}$, such that $\hat T_1$ corresponds to the strongest unit vote for the opponent party and $\hat T_{m^2}$ is the strongest  unit vote for the proponent. More precisely, the strongest unassigned opponent unit is $T_{i,j}\in U$ if $v_{i,j}\leq v_{k,l}$ for all $T_{k,l}\in U$, or equivalently it is $\hat T_\beta\in U$ if for all other  $\hat T_\gamma\in U$ one has $\beta<\gamma$. Conversely,  $\hat T_\beta$ for $\beta$ the largest  index in $U$ is the strongest unassigned proponent unit. 

Implementing a discretised version of the strategy outlined in \cite{FH}, our algorithm forms each district $D$ by iteratively adding the strongest the unassigned proponent unit followed by the strongest unassigned opponent units, until $P_D>P_{D_-}$. The algorithm then calculates the district vote $N_{D_k}$ and compares it to the vote threshold $w$. If $N_{D_k}>w$ and $P_D<P_{D_+}$ then the district is complete and the algorithm repeats this process to create the remaining districts, with the exception of the last district.  

It may be that in forming a given district, whilst the district satisfies $P_D>P_{D_-}$ the district vote is calculated to be less than the vote threshold. In this case the algorithm adds the strongest remaining  unassigned proponent units until $N_{D_k}>w$, and at each step checks that $P_D<P_{D_+}$. Once the district vote is sufficiently large, the district is complete. When the population limit is exceeded, $P_D>P_{D_+}$, our algorithm will remove the last unit added and try the next in the ordered list until it identifies an addition to the district that does not violate the population limit. For large vote thresholds $w$ or small population thresholds $t$, this districting algorithm may fail (i.e.~no district satisfies simultaneously eq.~(\ref{pop}) and (\ref{dkw})), but this is rarely a problem for percent-level $w$ and $t$.

 Finally,  the last district $D_n$ is identified with the remaining unassigned territories after the first ($n-1$) districts are constructed.     If the territory  is balanced, as we assume, then it is impossible for all districts to favour the proponent, and thus it is expected that for the final district $N_{D_n}<0$. Moreover, by design, the final district is primarily comprised of moderate voters.  The only requirement on the final district is that it satisfies  $P_{D-}\leq P_{D_n}\leq P_{D+}$ and this will commonly be the case for reasonable choices of $t$.
For a smaller population threshold, and thus a  more stringent requirement of population uniformity, the final district may  fail the population constraint. In this case, after constructing the final district the algorithm will make a number of amendments to the district compositions such that the populations are within the threshold. In the case that $P_{D_n} > P_{D+}$, the proponent-favouring territorial units on the exterior of the final district are transferred to adjacent districts. If $P_{D_n} < P_{D-}$, then the opponent-favouring territorial units in other districts and adjacent to the final district will be transferred to the final district.  

\begin{figure}[H]
	\centering
	\vspace{12mm}
	{\bf Sample Executions for Friedman-Holden Packing}\\[15pt]
		\hspace*{-13mm}\includegraphics[height=26.8mm]{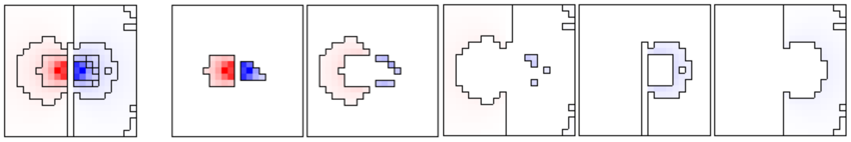}\\[10pt]
		\hspace*{-13mm}\includegraphics[height=27mm]{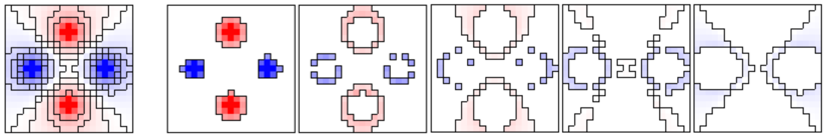}\\[5pt]
		\caption{Example districting results for benchmark model \#1 (top) \& \#4 (bottom) of Table \ref{tab0} with a Gaussian population distribution and a balanced vote. The gerrymander's party is coloured red, and in both cases wins the popular vote in three of the five districts.\label{Fig6a} }
	\end{figure}

\begin{figure}[H]
	\centering
	\hspace*{-10mm}
		\includegraphics[width=1.1\textwidth]{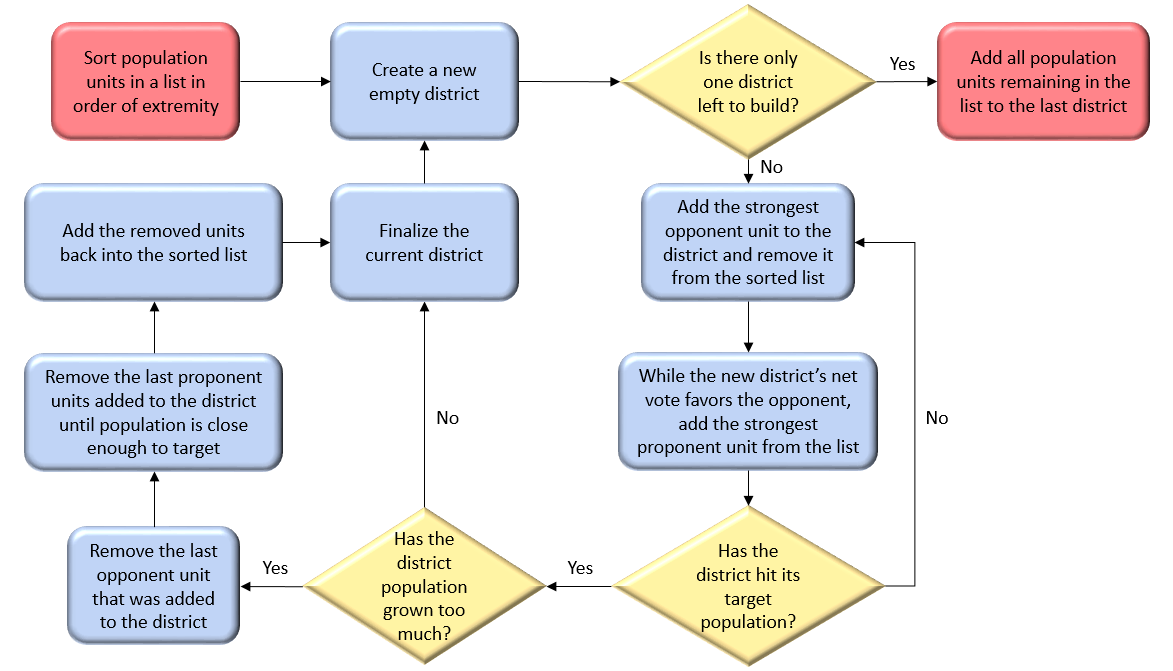}
		\caption{Flow chart of our implementation of the Friedman-Holden method. \label{FigA1} }
	\end{figure}

Example executions are shown in Figure \ref{Fig6a} and a flow chart illustrating the steps of this algorithm is presented in Figure \ref{FigA1}. Specifically, we show the output of our implementation of the algorithm of \cite{FH}, as described above, partitioning a $21\times21$ lattice territory into  5 districts for benchmark models \#1 and \#4 (defined in Table \ref{tab0}). The intensity of the colour indicates the partisan extremity of a given territorial unit  and the black lines indicate divides between districts. The left most panel illustrates the whole territories, whilst the panels to the right show, in order, the composition of Districts 1 to 5 in order of construction. Observe that the district compositions output by this algorithm are all typically disconnected, and this shall be quantified  through Monte Carlo  studies shortly.

\subsection{Impact of Friedman and Holden Packing Method}
\label{secbbb}

Before examining whether the districts constructed emulating the Friedman and Holden method are connected we  shall first look to quantify what degree of advantage this gerrymandering procedure gives to the proponent by comparing to a non-partisan model of districting.  To assess the impact of gerrymandering we construct a voter model on a  $21\times21$ lattice territory  $S$ where the population distribution is quasi-Gaussian as in Section \ref{2.2}, the source points follow the benchmarks of Table \ref{tab0}, and the total population\footnote{The typical population of U.S. congressional districts is 700,000; our results can be rescaled accordingly if desired.} is fixed to be $P_S\approx4700$ (up to $0.5\%$ fluctuations). 
 Since the walker function introduces random fluctuation in the approximately Gaussian population distribution, each run returns a different redistricting. Thus we can assess the impact of gerrymandering for a given set of source points via a Monte Carlo approach using multiple runs of the algorithm. 

	\vspace{-3mm}
\afterpage{\clearpage}\clearpage

 \begin{wrapfigure}{R}{0.45\textwidth}
  \ffigbox[0.9\FBwidth]
  {   \caption{A fair symmetric districting of a $21\times 21$ lattice. 
  \label{Fig7}} }
  {    \includegraphics[width=0.8\textwidth]{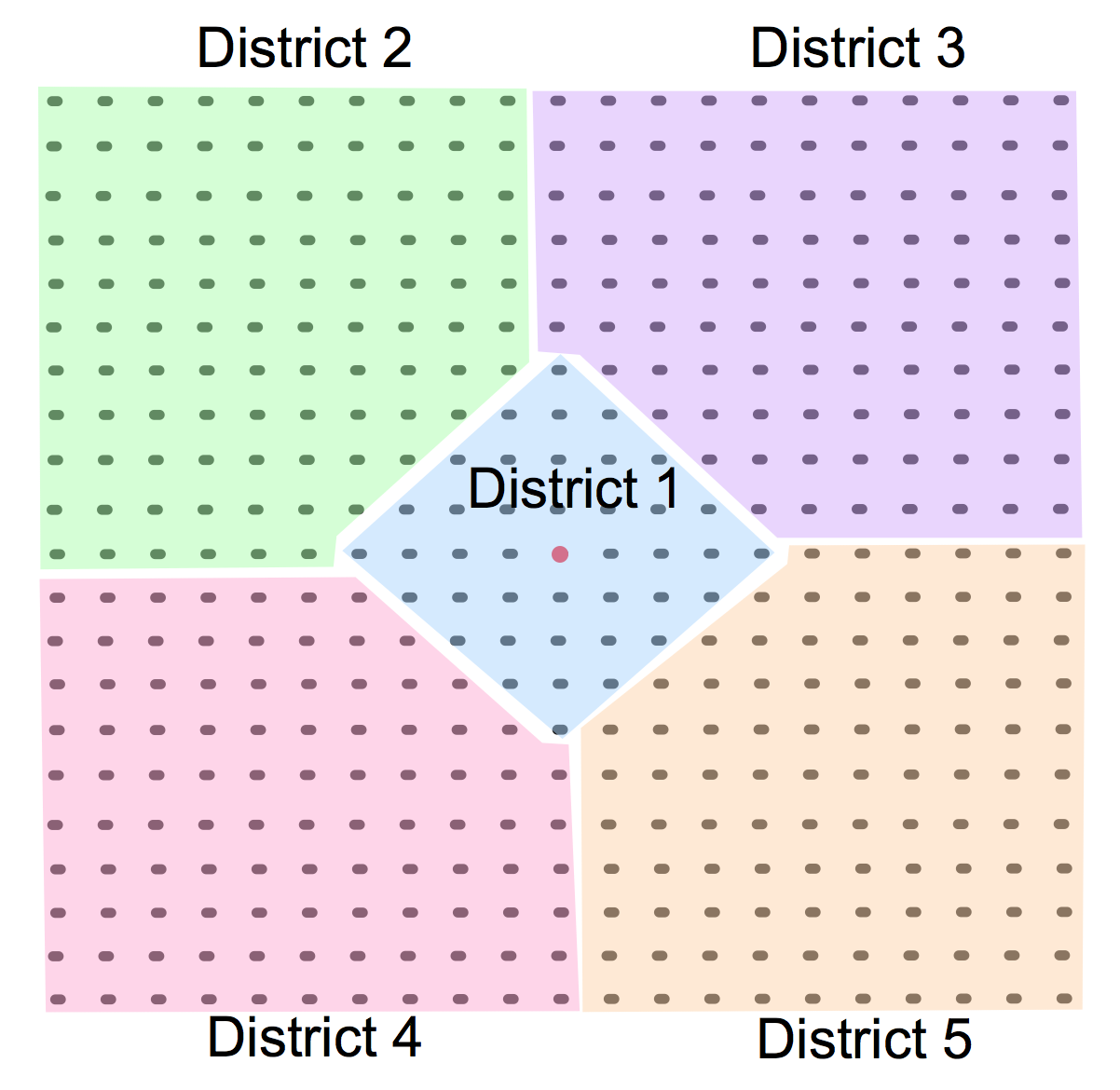}  }
\end{wrapfigure}

 Specifically, we generate a set of distinct redistricting plans on 30 different $21\times 21$ lattice population models with balanced votes $N_S\approx0$  for each of the four benchmark source point placements (as shown in Figures \ref{Fig4} \& \ref{Fig4b}), and partition the territory into five districts (i.e.~take $n=5$). For each district $D_k$,  for $1\leq k\leq5$, we calculate the average district vote $N_{D_k}$ and population $P_{D_k}$,  and  the average number of district wins for the proponent $\#_{\rm win }$. In Table \ref{tab1} we show the average net vote $N_{D_k}$  in each district (where the districts are enumerated by order of construction) following redistricting via FH packing and averaging over 30 runs:
 

\begin{table}[H]
\begin{center}
\def\str{\vrule height12pt width0pt depth7pt}
\begin{tabular}{| c | c | c | c | c | c  | c | c | c | c | c | c | c | c | c  | c | c |  c | c | c | }
    \hline\str
Model  & $P_{D1}$ &$N_{D1}$ & $P_{D2}$ &$N_{D2}$ &$P_{D3}$ & $N_{D3}$ & $P_{D4}$ &$N_{D4}$ & $P_{D5}$ &$N_{D5}$ & $\#_{\rm win }$ 
   \\  
\hline
 \#1 &    902 & 43.7 & 970 & 5.8 & 978 & 1.5 & 946 & -34.6 & 910 & -16.3 & 3   \\    
 \#2    & 898   & 60.6       &  903     &  16.0   &  955      &   9.5  &  950  & -27.2  & 1001  &  58.8   &  3 \\
 \#3       & 897   & 82.1    & 899   &  37.2  &   906      &   29.4    & 950   &   -21.6    & 1059   &  -127.1  &   3     \\
 \#4    & 901   & 83.6    & 898   &  15.4  &   939      &   9.8    & 945   & 69.6    & 1022   &  39.8  &   3        \\
\hline
\end{tabular}
\caption{Average net vote per district following redistricting via the FH method. \label{tab1}}
\end{center}
\end{table}


It is insightful to compare these results to some ``fair'' partitions of the population. Since the population approximates a two dimensional Gaussian distribution, any spherically symmetric partition of the population which does not take into account the voter distribution can be considered a non-partisan districting. We shall construct a non-partisan $n=5$ districting by allocating the central cells to District 1 and then symmetrically partitioning the set of territorial units not assigned to District 1 to create Districts 2-5.

We choose to partition Districts 2-5 by simply drawing the district lines along the horizontal and vertical mid axes, and assigning the territorial units on the  borders to the adjacent district in the clockwise direction. The size of District 1 is fixed such that the 5 territories have approximately equal populations,  thus it is determined by  the Gaussian spread and population threshold $t$.   The resulting districts are illustrated in Figure \ref{Fig7}.

To assess the impact of gerrymandering we compare predicted results for the case of aggressively gerrymandered districts (as determined by our algorithmic implementation of the FH packing strategy) to the vote for the symmetric districts outlined above for balanced territories. Generating 30 population distributions on $S$, we calculate both the average net vote per district $N_{D_k}$,  and the average number of district wins for the proponent $\#_{\rm win }$, as shown in Table \ref{tab2}.

\begin{table}[H]
\begin{center}
\def\str{\vrule height12pt width0pt depth7pt}
\begin{tabular}{| c | c | c | c | c | c  | c | c | c | c | c | c | c | c | c  | c | c |  c | c | c | }
    \hline\str
Model &  $N_{D1}$ & $N_{D2}$ & $N_{D3}$ & $N_{D4}$ & $N_{D5}$  & $\#_{\rm win }$ 
   \\  
\hline
 \#1    & 0.1   & 35.1      &  31.5      & -31.5  & -35.1    &  2.44 \\
%
 \#2    & -0.1   &  153.6       &   124.0        &  -124.1    &  -153.5    &    2.56 \\
%
 \#3         & -0.1   & -311.1    & 49.3   &  -49.0  &   310.8   &   2.46  \\
%
 \#4       & 0.0   & -45.2    & 44.8   &  44.6  &   -44.9 &    2.42  \\
%
\hline
\end{tabular}
\caption{Average net vote per district for idealized non-partisan symmetric districts.\label{tab2}}
\end{center}
\end{table}

Through the comparison of Tables \ref{tab1} \& \ref{tab2} it is clear that FH packing significantly impacts the electoral outcome, skewing the predicted number of district wins  $\#_{\rm win }$  in favour of the proponent party. However, as we show next, the districts constructed via FH packing are generically disconnected and therefore typically legally prohibited.

\subsection{Connectivity of Friedman and Holden Districts}
\label{3.1}

Since the Friedman-Holden method includes no spatial data,  one might expect disconnected districts. In what follows we shall quantify the failure of the FH packing method to produce connected (and thus legal) districts. To this end we shall  take the districts constructed in Table \ref{tab1}, output the assignments of the territorial units, and identifying the number of connected components in each district.

To show that the districts constructed via FH packing are highly disconnected we take the outputs of the 30 runs of FH packing on $S$ generated previously and calculate the  mean number of connected components for each district (labelled D1 to D5) and the mean over all districts for the 4 benchmark source point models. The results are displayed in Table \ref{tab3} and Figure \ref{Fig6a} shows one (of the 30) district composition output by the algorithm for each benchmark model, where one can see that the districts are highly disconnected.

\begin{table}[H]
\begin{center}
\def\str{\vrule height12pt width0pt depth7pt}
\begin{tabular}{| c | c | c | c | c  | c | c | c  | c |}
    \hline\str
Model   & D1 & D2 & D3 & D4 & D5 & Average  \\  
\hline
 \#1        & 2.4       & 4.9        & 8.7    & 1.1       & 1.4        & 3.7         \\ 
\#2        & 2.0       &  8.7        & 9.0    & 9.1       & 1.0        & 6.0         \\ 
 \#3         & 4.0   & 18.2       & 14.4      & 12.5    & 1.0     & 10.2     \\ 
 \#4          & 4.0   & 16.6       & 24.1      & 26.1    & 6.9     & 15.5       
         \\ 
\hline
\end{tabular}
\vspace{-3mm}
\caption{Number of connected components for districts created by our algorithmic  implementation of the FH packing strategy.\label{tab3}}
\end{center}
\end{table}
\vspace{-3mm}

	
Under the above settings we find the strategy  of \cite{FH} typically outputs districts with $\mathcal{O}(10)$ components. Importantly, disconnected districts are legally prohibited. Thus, while the Friedman-Holden method is simple and easy to implement, it is too simplistic to give a realistic model of redistricting and makes it clear that it is critical that algorithmic approaches  take into account the spatial distribution of voters. We note also that \cite{PT2} have given complementary arguments against ignoring spatial data whilst gerrymandering from an axiomatic perspective.

	\section{Packing with Spatial Restriction} 
	\label{sec4}

We dedicate this section to introducing two novel algorithms that gerrymander voter distributions on a lattice in a manner that strongly favours the proponent and gives mostly connected districts:
 ``Spatially Restricted Friedman-Holden'' (SRFH) Packing and   ``Saturation'' Packing. 
 In SRFH packing the gerrymander aims to guarantee wins in the majority of districts, but lose the later constructed (moderate) districts, whereas the saturation strategy relies on constructing districts to be discarded with significant opponent biases.
  In Section \ref{sec4b} we introduce one further new strategy based on genetic algorithms. Python codes for each of our algorithms are provided online.

\vspace{-1mm}
\subsection{Spatially Restricted Friedman-Holden Packing}
\label{srfh}
\vspace{-1mm}

	We shall introduce here the  Spatially Restricted Friedman-Holden (SRFH) Packing algorithm, which is a modification to the FH packing approach that allows the inclusion of spatial data.  The SRFH districting algorithm forms each district by first including the strongest opponent and proponent unassigned units, as well as a path of territorial units between them (which necessarily includes moderate voters). The algorithm then successively adds highly polarized units of both parties which are adjacent to the forming district, until it satisfies the population requirement and the net vote favours the proponent party, i.e.~until the district satisfies eq.~(\ref{pop}) and eq.~(\ref{dkw}).
	
\begin{figure}[t!]
		\includegraphics[width=0.85\textwidth]{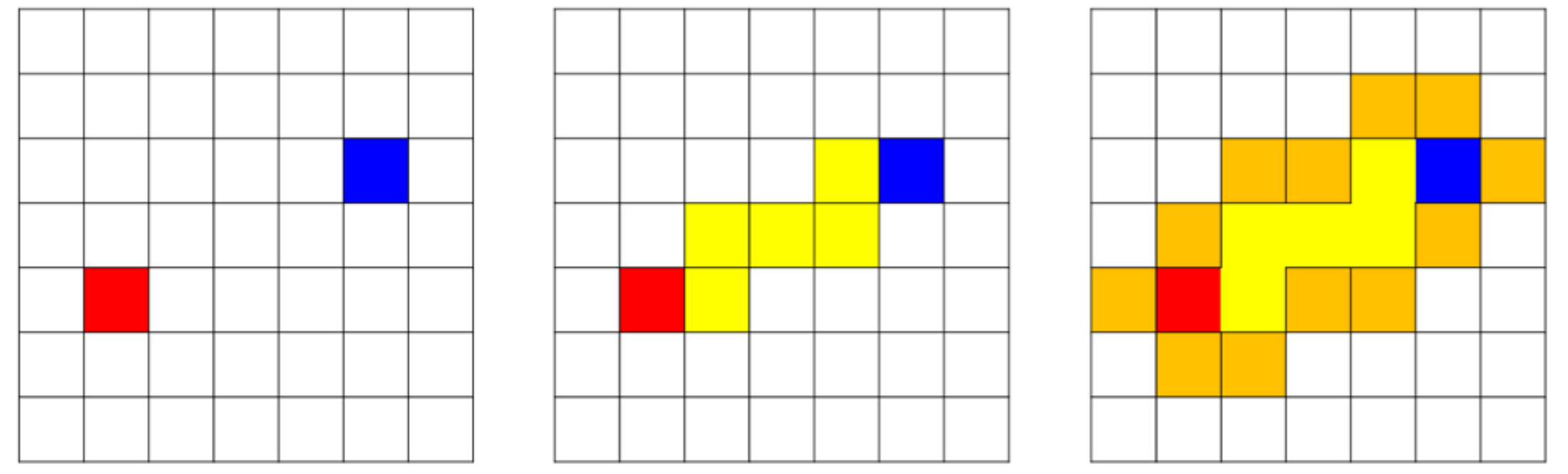}\\
		\caption{\label{path} Left: An example of the most extreme proponent and opponent units. Centre: the shortest path through unassigned territory between the extreme endpoints. Right: the set of territorial units (orange) considered for addition to the district. }
	\end{figure}

	 We first discuss the general construction of districts, with the exception of the final district. Suppose there are $x$ unassigned territorial  units, then similar to the FH algorithm the first step is to sort, via quicksort, the unassigned units in order of decreasing net vote $v_{i,j}$ and relabel the ordered set $\{\hat{T}_1,\hat{T}_2 \cdots \hat{T}_x\}$. To form a new district the algorithm includes the unassigned units  $\hat T_1$ and $\hat T_x$ in the district and the shortest path between them that avoids all already assigned territorial  units, as illustrated in Figure \ref{path}. We use a standard {\em Grassfire} algorithm \cite{grassfire} to remove assigned units when determining the shortest path. If a given path between $\hat T_1$ and $\hat T_x$ exceeds the target population $P_{D\pm}$, then the next shortest path can be selected instead, however this is typically not an issue.

	While $P_{D_k}<P_{D-}$ the algorithm will repeatedly add to the district the most extreme territorial unit that is unassigned and adjacent to it. If the net vote of the district does not sufficiently favour the proponent party, as determined by a vote threshold  parameter $w$,  then the algorithm will successively add proponent units to the district. Once the net vote favours the proponent, the algorithm adds opponent units to balance the vote. After adding each unit to a given district, the algorithm calculates the current district population $P_D$ and fixes the district when $P_D>P_{D-}$. It then checks that $P_D<P_{D+}$, if this check fails, the most recent territorial unit added is replaced with an alternative adjacent unit, until the above is satisfied. After a connected set is validated as a district, the algorithm recurs on the remaining sorted list of territorial units to create another district. The algorithm will typically produce districts in order of decreasing $N_{D_k}$, with the proponent vote being stronger than the opponent vote. The remaining unassigned units after the construction of $n-1$ districts are assigned to the final district, and thus the last district is typically disconnected.  Figure \ref{FigA2} provides a flow chart illustrating the steps of this algorithm. We will quantify the typically number of connected components in Section \ref{3.3FH}.

\subsection{A Saturation Packing Algorithm}
\label{sat}

In what follows we shall introduce an alternative packing algorithm, we call the Saturation Packing algorithm, which implements the classic strategy of simply packing extreme opponent voters into a single compact district. The Saturation Packing algorithm builds on the establish algorithmic framework of the SRFH Packing method, however prior to constructing any districts, each territorial unit is assigned a \textit{priority} value based on its net vote and its average distance from proponent source points.
\begin{definition}
	For a territory $S$ with $\alpha$ proponent source points $E_1, \ldots, E_\alpha$, we define the \textit{\bf priority} $z_{i,j}$ of a territorial  unit $T_{i,j}$ to be 
	\begin{equation}
	z_{i,j} := -v_{i,j} \cdot \frac{1}{\alpha}\sum_{i=1}^{\alpha}d\left(E_i, T_{i,j}\right)~.
	\end{equation}
\end{definition}

Once all priorities are assigned, the list of territorial units is sorted in order of decreasing priority and the territorial unit with the greatest negative priority at the beginning of the sorted list is added to the collection $D_1$, that will eventually form the first district. Then, the algorithm searches over all territorial units adjacent to $D_1$ and adds the next highest negative priority unit to $D_1$. This last process is repeated until $P_{D_1} > P_{D-}$. After District 1 is constructed via the Saturation Packing algorithm, all subsequent districts are created using the SRFH Packing algorithm outlined in the previous section. A flow chat for this algorithm can be seen in Figure \ref{FigA3}.

\subsection{Example Executions of   Saturation Packing and Friedman-Holden Packing }
\label{3.3FH}

\begin{figure}[H]
	\centering
	\vspace{12mm}
	{\bf Sample Executions for Spatially Restricted Friedman-Holden Packing}\\[15pt]
		\hspace*{-13mm}\includegraphics[height=26.8mm]{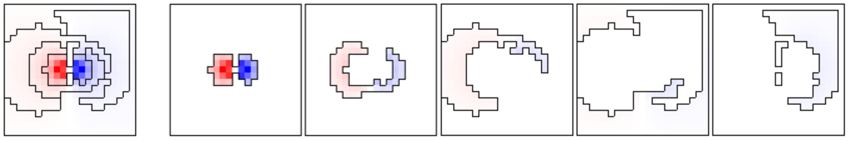}\\[10pt]
		\hspace*{-13mm}\includegraphics[height=27mm]{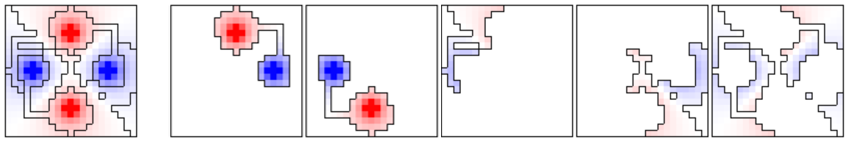}\\[5pt]
\caption{The Spatially Restricted Friedman-Holden Packing algorithm applied to benchmark voter model \#1 (top) \&  model \#4 (bottom)   of Table \ref{tab0}, analogous to Figure \ref{Fig6a}, in both red wins 3 districts.
 \label{Fig6b} }
	\end{figure}

\begin{figure}[H]
	\centering
	\hspace*{-10mm}
		\includegraphics[width=1.1\textwidth]{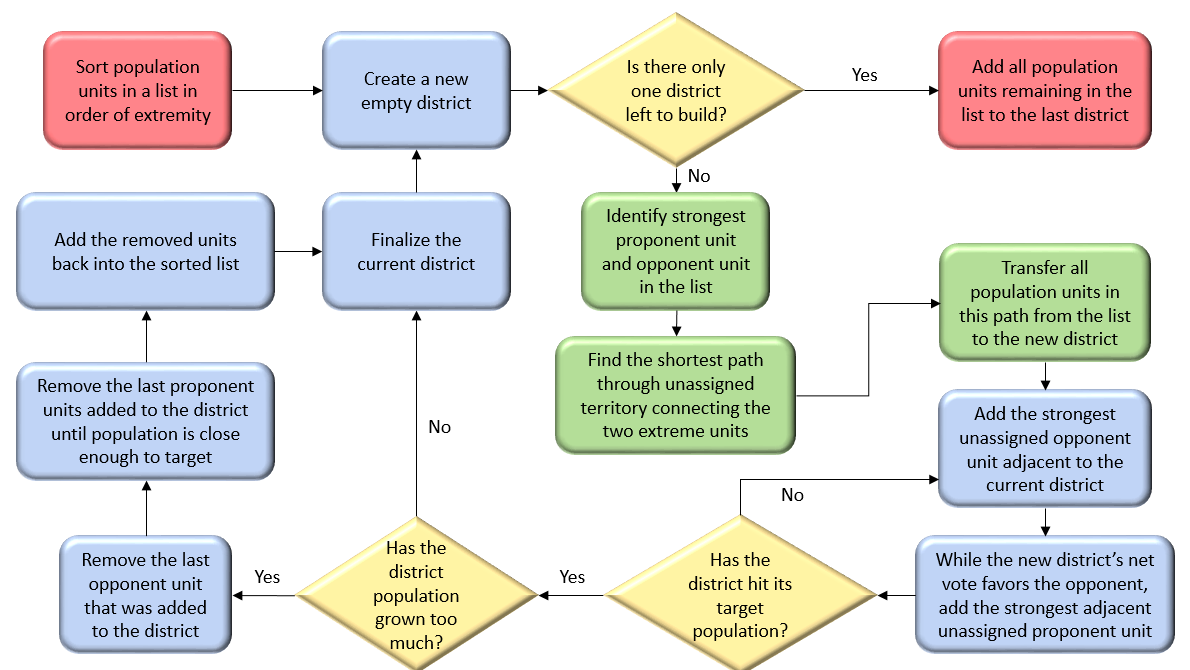}
		\caption{Flow chat of Spatial Restricted Friedman-Holden Packing of Section \ref{srfh}. \label{FigA2} }
	\end{figure}

\begin{figure}[H]
	\centering
		\vspace{12mm}
	{\bf Sample Executions for Saturation Packing}\\[15pt]
		\hspace*{-13mm}\includegraphics[height=27mm]{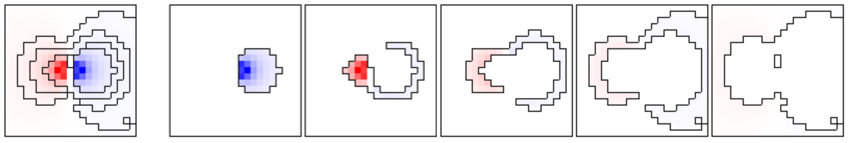}\\[10pt]
		\hspace*{-13mm}\includegraphics[height=27mm]{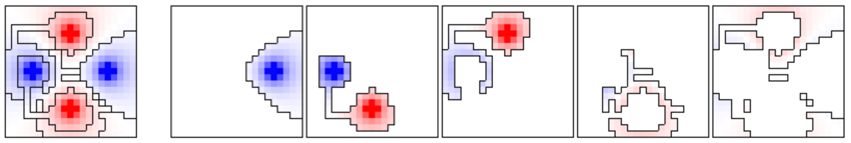}\\[15pt]
		\caption{Saturation Packing  applied to the benchmark voter model \#1 (top) and  model \#4 (bottom)  of Table \ref{tab0}, analogous to Figures \ref{Fig6a} \& \ref{Fig6b}, in both red wins 4 districts.
		\label{Fig6c}}
	\end{figure}

\begin{figure}[H]
	\centering
		\hspace*{-10mm}
	\includegraphics[width=1.1\textwidth]{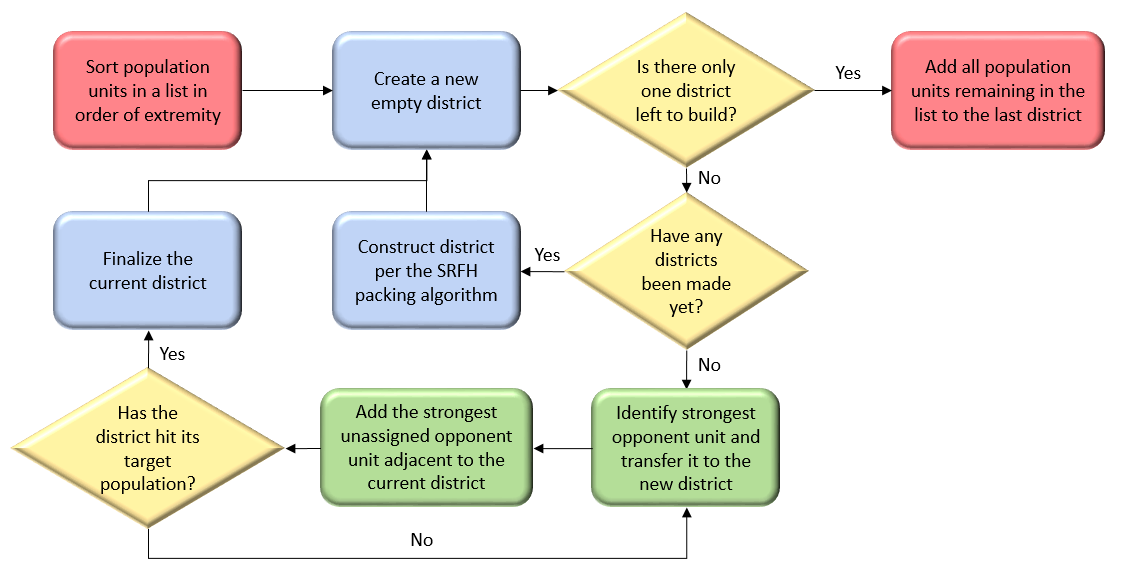}
		\caption{Flow chat of Saturation Packing detailed in Section \ref{sat}. \label{FigA3} }
	\end{figure}

In Section \ref{sec5} we will use the two algorithms developed in this section to undertake Monte Carlo studies of aspects of gerrymandering. First, however, it is insightful to consider a number of individual examples, and in particular to view the typical districts which each algorithm produces for a given voter distribution, analogous to Figure \ref{Fig6a}.
 Figures \ref{Fig6b} and \ref{Fig6c} present example outputs of, respectively, the SRFH Packing and Saturation Packing  strategies in which we partition a $21 \times 21$ lattice into five districts for the benchmark voter models \#1 and \#4 of Table \ref{tab0}.

\afterpage{\clearpage}\clearpage

\afterpage{\clearpage}\clearpage


	Observe that, by design, with the exception of District 5 the districts created are connected, which is a notable improvement on the original strategy of \cite{FH}  which typically produced an average of $\mathcal{O}(10)$ component for each district. Following similar methodology to Table \ref {tab3} with 30 trails, in Table \ref{tab3b} we quantify the number of connected component in the final district (``D5 \#'') for the  Saturation Packing and SRFH-Packing, as well as stating the average number of connected components for all districts (``average'') to allow comparison with Table \ref {tab3}. 
\begin{table}[H]
\begin{center}
\def\str{\vrule height10pt width0pt depth7pt}
\begin{tabular}{| c | c | c | c | c  | c | c | c  | c |}
    \hline\str
Model   & SRFH: D5 \# & SRFH: average & Saturation: D5 \#  & Saturation: average\\  
\hline
 \#1         & 3.6   & 1.5    & 2.8   & 1.4        \\ 
 \#2          & 4.4   & 1.7    & 4  & 1.6          \\ 
 \#3         & 5.7   & 1.9       & 6.0   & 2.0    \\ 
 \#4          & 8.2   & 2.4      &  3.9   & 1.6 \\
\hline Average  & 5.5 & 1.9 & 4.2 & 1.7
         \\ 
\hline
\end{tabular}
\vspace{-2mm}
\caption{Average number of connected components for District 5, and average number of connected components over all five districts, for  the Saturation and SRFH algorithms.\label{tab3b}} 
\end{center}
\end{table}
\vspace{-2mm}
As shown in Section \ref{3.1} using the basic FH Packing method, which does not account for the spatial distribution of voters, the average number of connected components is $\mathcal{O}(10)$ components (cf.~Table \ref {tab3}), whereas for our algorithms with spatial restrictions the average is $\mathcal{O}(1)$. With so few components one can potentially achieve connectivity in post-processing with a small number of swaps. Moreover, in a real world environment one could also possibly exploit geographical features or utilize the empty area external to territory to arrange for the final district to be connected. Studies on the effectiveness of these different districting strategies shall be presented in Section \ref{sec5}.

	\section{Genetic Gerrymandering Algorithm} 
	\label{sec4b}

In Section \ref{sec4} we advanced on the basic strategy of \cite{FH} by including spatial restrictions through what we called the Saturation and SRFH packing algorithms. However, those methods both rely on assigning territorial units to districts in manner that satisfies local requirements without reference to the global assignments and thus typically the final districts are disconnected. Whilst one can potentially `fix' the final district following the completion of the algorithm, this is somewhat unsatisfactory.  
In order to resolve this issue that the final district remains disconnected, we introduce here a new class of gerrymandering algorithm which we call Genetic Gerrymandering (GG), based on the general ideas of genetic algorithm approaches. Rather than constructing districts by selecting particular territorial units, this genetic algorithm takes a starting configuration of districts and evolves it over a number of iterations towards some goal. 
From some arbitrary initial `seed' set of districts the algorithm generates a number of mildly altered variants (mutations), it then calculates a fitness index for each of the sets of districts and retains the district sets with the highest index values, before repeating the process. After several iterations the system converges on a particular outcome. 

One distinct benefit of this procedure is that since the initial seed is a set of connected districts, provided that connectivity is respected in subsequent mutations, then the final set of gerrymandered districts will also be connected. 
To our knowledge genetic algorithms have not been applied to the question of how to optimally gerrymander a territory to provide highly partisan outcomes. There has, however, been a good deal of work on implementing fair districting using genetic algorithms, e.g.~\citep{Forman,Bacao,Chou2,Bard,Vanneschi}.

\subsection{The Gerrymandering Index}

In order to gerrymander the districts, one needs to construct a fitness index which favours the proponent party. Although there is no unique way to construct this index, any good fitness index for gerrymandering should take into account the number of districts which are winnable for the gerrymanders' party and the population spread. The latter  is taken into consideration to ensure that the districts have approximately equal populations, in order to make the districts legally valid. The fitness function we construct depends on these two qualities and it allows the algorithm to skew the populations if it provides an advantage to the proponent party.

Let $W$ be the number of districts which are winnable by the proponent, given by the number of districts with $N_{D} > w$,  plus half the contended districts (those with $\left|N_{D}\right| < w$)
\begin{equation}
W=\sum_i \left[\Theta(N_{D_k}-w) +\frac{1}{2}\left(\sum_i \Theta(N_{D_k})-\sum_i \Theta(N_{D_k}-w)\right)\right]~,
\label{WW}\end{equation}
where $\Theta$ is the Heaviside $\Theta$ function.
 We also define the following population measure
\begin{equation}
g :=  \frac{1}{P_S}\left(\max\limits_{1\leq i \leq n}\left(P_{D_k}\right) - \min\limits_{1\leq i \leq n}\left(P_{D_k}\right)\right)~.
\label{fP}
\end{equation}
The quantity $W$ and the measure $g$  characterize the salient properties of a given district plan for the purposes of gerrymandering, and using these we define the following gerrymandering fitness index, given by the ratio of districts won minus the population spread:
\begin{equation}
G := \frac{W}{n}-g~,
\label{gggg}
\end{equation}
The index $G$ takes values in the interval $\left[0, 1\right]$, with the most desirable outcome for the gerrymanderer occurring for those sets of districts with the highest fitness index. 

This index $G$ we consider is fairly simple, defined by just two parameters, and one could certainly construct more elaborate indexes, but this simple form is sufficient to explore the prospects of gerrymandering  a given territory.  However, identifying highly effective indexes for gerrymandering would be  an interesting question for future work.

\subsection{Random Seed}

Having quantified which sets of districts are preferred through the gerrymandering index $G$, the other components of the genetic algorithm are the definition of a set of seed districts for the genetic gerrymandering algorithm, being a random starting configurations of districts, and the mutation procedure which generates the iterations. 

Our starting point for the genetic gerrymandering algorithm  is an arbitrary partition of the territory into $n$ connected districts which is referred to as the seed. We construct this random seed as follows: for a partition of the territories into $n$ districts we randomly select $n$ territorial units, label these $\tilde T_\beta$ for $\beta=1,\cdots n$, and perform a flood fill \citep{Smith,Glassner} around these units.  Via the flood fill algorithm each of the territorial units $T_k\in S$  is assigned to the nearest unit in $\{\tilde T_i\}$, calculated using the distance function defined in eq.~(\ref{eqd}). The districts in the seed are defined  as follows
\begin{equation}
D_k:=\{T_k \in S ~|~d(T_k-\tilde T_i)\leq d(T_j-\tilde T_k)~{\rm for}~i\neq j\} \qquad (i=1,\cdots n)~.
\end{equation}
One should note that via this definition of $D_k$ equidistant points would not be assigned to a district. This is a minor technicality and in the algorithmic definition these unassigned units are simply assigned to the set carrying the lowest numerical label $i$.

  The flood-fill random districting guarantees connected districts that are typically convex and compact. Although this method does not enforce population equality, the random districtings start out highly equitable from a geometric standpoint, and the district populations tend to balance out as they evolve due to the definition of the index $G$.

\subsection{Mutation Procedure}

The final component of the genetic gerrymandering algorithm is to randomly alter the districts, identify the mutated district which maximizes $G$, and iterate this procedure. Our mutation procedure swaps certain units between adjacent districts on each iteration  in a manner that preserves connectedness of each district and favours balanced populations.
\begin{definition}
	For a territorial unit $T$ in district $D$, a district $D'$ (for $D\neq D'$) is said to be an {\bf adjacent district} to $T$ if there exists a $T'\in D'$ which is an adjacent unit to $T$.
\end{definition}
To describe a territorial units propensity to switch to a different adjacent district during the mutation procedure we define the following
\begin{definition}
	The {\bf instability} $L_{i,j}$ of  a territorial unit $T_{i,j}$ in district $D$ is zero if $D-T_{i,j}$ is not connected. Otherwise, $L_{i,j} := \max\left(P_D - P_{D'}\right)$ over all districts $D'$ adjacent to $T_{i,j}$.
\end{definition}
From the above definition, if a given territorial unit $T_{i,j}$ lies in the same district as each of its adjacent territorial unit, then $L_{i,j} = 0$ and $L_{i,j} \neq0$ occurs only for territorial units on a border between districts. To implement the mutation process, we first determine the instability $L_{i,j}$ for each $T_{i,j}\in S$ and we define $M:={\rm max}[L_{i,j}]$ over all $T_{i,j}\in S$.

\newpage
For all $T_{i,j}\in S$ with $L_{i,j} > 0$, a mutation on $T$ is performed with probability $L_{i,j}/M$, removing $T_{i,j}$ from its old district and adding $T_{i,j}$ to the lowest-population district adjacent to $T_{i,j}$. This transfer of territorial units will always make the district populations more balanced and thus the voting districts become significantly more equitable rather quickly (in just a few generations).

In summary, a set of $N$ seed districts forms generation 1 and our algorithm then performs the following iterative procedure for generation $i$: it identifies the $\sqrt{N}$ cases (rounding appropriately) with the highest $G$ indices and to these $\sqrt{N}$ cases it applies the mutation procedure $\sqrt{N}$ times. Because of the probabilistic nature of the mutations this produces $N$ different offspring, which comprise generation $i+1$. With each iteration the set of districts evolve towards a local maximum of $G$. After a preset set number of iterations the algorithm outputs the set of districts with the best $G$ index from the final generation. Typically, after $\mathcal{O}(10)$ iterations the algorithm converges to sets with $G\simeq1$. We provide a flow chart illustration of this algorithm  in Figure \ref{FigA4}.

\subsection{Sample Executions of  Genetic Gerrymandering}
\label{s4d}

\begin{figure}[H]
	\centering
	\advance\rightskip-2cm
	\vspace{12mm}
	{\bf Sample Executions for Genetic Gerrymandering}\\[15pt]
		\hspace*{-15mm}\includegraphics[height=26.8mm]{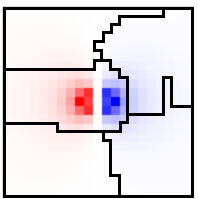}
		~~
		\includegraphics[height=26.8mm]{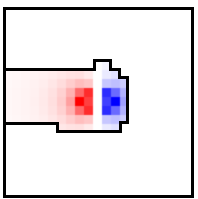}
		\hspace*{-2.5mm}
		\includegraphics[height=26.8mm]{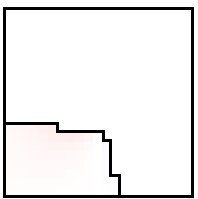}
		\hspace*{-2.5mm}
		\includegraphics[height=26.8mm]{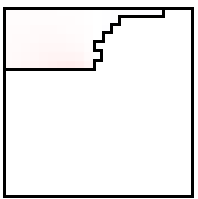}
		\hspace*{-2.5mm}
		\includegraphics[height=26.8mm]{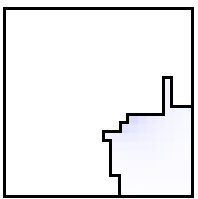}
		\hspace*{-2.5mm}
		\includegraphics[height=26.8mm]{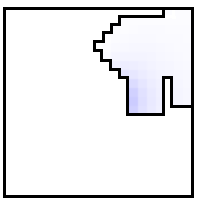}
		\\[10pt]
		\hspace*{-15mm}\includegraphics[height=26.8mm,angle =90]{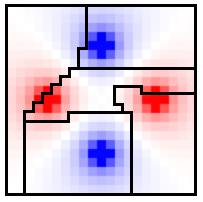}
		~~
		\includegraphics[height=26.8mm,angle =90]{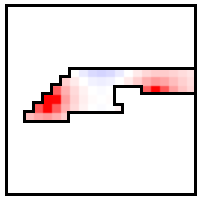}
		\hspace*{-2.5mm}
		\includegraphics[height=26.8mm,angle =90]{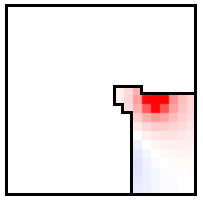}
		\hspace*{-2.5mm}
		\includegraphics[height=26.8mm,angle =90]{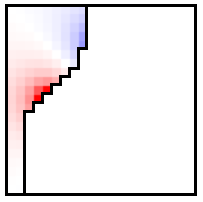}
		\hspace*{-2.5mm}
		\includegraphics[height=26.8mm,angle =90]{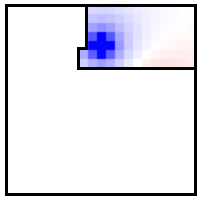}
		\hspace*{-2.5mm}
		\includegraphics[height=26.8mm,angle =90]{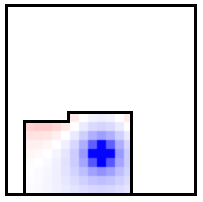}\\[8pt]
		\caption{Genetic Gerrymandering  applied to benchmark model \#1 (top) and model  \#4 (bottom) of Table \ref{tab0}, analogous to Figures \ref{Fig6a}, \ref{Fig6b} \& \ref{Fig6c}, in both red wins 3 districts.
\label{Fig6d}    } 
	\end{figure}

\begin{figure}[H]
	\centering
		\hspace*{-10mm}
		\includegraphics[width=1.1\textwidth]{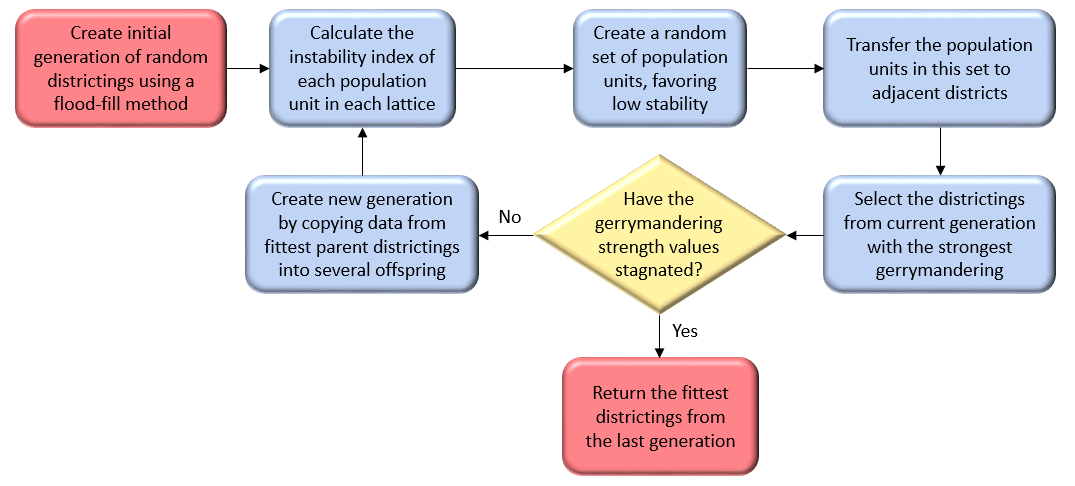}
		\caption{Flow chat of Genetic Gerrymandering algorithm of Section \ref{sec4b}. \label{FigA4} }
	\end{figure}

For our example run we construct $N= 100$ random seed districtings for the first generation and the algorithm selects the 10 with highest $G$ index to mutate. At each stage the algorithm runs the mutation method $10$ times on each parent, producing $10\cdot10 = 100$ offspring. This evolution is iterated over 30 generations, and the final districting output is the set in the last generation with the highest $G$ index. In Figure \ref{Fig6d} we present a sample executions of the genetic gerrymandering algorithm applied to the benchmark models  \#1 and \#4 of Table \ref{tab0}, in analogy to Figures  \ref{Fig6a}, \ref{Fig6b} \& \ref{Fig6c}.

\subsection{Modified GG for Fair Redistricting}

Before closing this section we highlight that the genetic gerrymandering algorithm developed above can be adapted to construct fair districtings simply by changing the fitness index. As mentioned above, the use of genetic algorithms to fair districting has been previously studied in the literature, for instance \citep{Forman,Bacao,Chou2,Bard,Vanneschi} and, indeed, many of the existing algorithms for fair districting are far more sophisticated than the one we discuss  below. However, the genetic heuristic and mutation procedure developed in our genetic gerrymandering algorithm is distinct and original, and it is interesting that, at least in our simplified lattice models, fair districts can be readily constructed from the simple and intuitive criteria we stipulate below.

 While fairness is a subjective concept, one can construct a reasonable fitness index via a heuristic formula that takes into account:
 \begin{itemize}
\item[{\em i)}] The district population balance;
\item[{\em ii)}]  Accuracy of state representation in Congress;
\item[{\em iii)}]   Representation accuracy in each district.
 \end{itemize}

\afterpage{\clearpage}\clearpage
\newpage

  We quantify these three characteristics, respectively, via fitness functions $f_P$, $f_S$, $f_D$:
\begin{equation}
\begin{aligned}
f_P &:= 1 - g,\\[5pt]
f_S &:= 1 - \left|\frac{1}{2}\left(1 + \frac{N_S}{P_S}\right) - \frac{W}{n}\right|,\\[5pt]
f_D &:= \frac{1}{n}\sum_{i=1}^{n}\frac{Q_{D_k}}{P_{D_k}},
\end{aligned}
\end{equation}
where $W$ and $g$ are defined in eq.~(\ref{WW})  \& (\ref{fP}), and we denote by $Q_D$ a measure of the proportion of the population in a given district $D=\cup_{(i,j)\in I} T_{i,j}$ (for an index set $I$) whose vote preference $v_{i,j}$ aligns with the net district vote $N_D$, defined by:
\begin{equation}
Q_D:= \sum_{(i, j) \in I} P_{i,j}\cdot\frac{1}{2}\left(1 + \text{sign}[v_{i,j}]\cdot  \text{sign}[N_D]\right)~.
\end{equation}
Note also that whereas $g$ measured the population spread, here $f_P$ characterises the population balance.

\begin{figure}[b!]
	\includegraphics[width=0.8\textwidth]{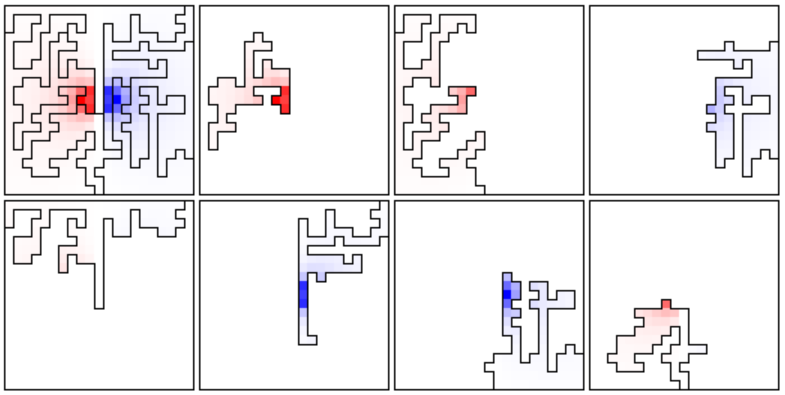}
	\caption{Construction of 7 fair districts via the Modified GG for Fair Redistricting applied to a $21\times21$  territory with balanced vote and two source points (with placement as in model \#1 of Table \ref{tab0}). Presentation is analogous to Figure \ref{Fig6a}. 
	 \label{Fig6e} }
\end{figure}

Using the  partial fitness functions, we define the overall ``fair'' fitness $F$ to be
\begin{equation}
F := \frac{2}{5}\left(f_P + f_S + f_D - \frac{1}{2}\right)~,
\label{ffff}
\end{equation}
whose values lie in the interval $\left[0, 1\right]$. Districts that are deemed ``fairest'' should have the highest fitness index. Typically for our trails we have found that the final set of fair districts tends to have fitness index values in excess of $F > 0.9$. 

\newpage

We will refer to the Genetic Gerrymandering (GG) algorithm equipped with the index $F$ (rather than $G$) as the ``Modified-GG for Fair Redistricting''. Following the same methodology of Section \ref{s4d} we execute an example run using $N=100$ seed configurations and taking 30 iterations.  The results of this example run are shown in Figure \ref{Fig6e} for seven districts. In particular, by design, the equitable district configurations produced by the Modified GG for Fair Redistricting support a $50\%-50\%$ vote split for balanced territories, accurately representing voter preference across the whole territory. Moreover, a majority of the districts avoid including oppositely polarized voters, ensuring that each elected district representative reflects the values of their constituents.

\section{Results} 
\label{sec5}
	
Through the algorithms developed here we shall quantify some general statements about gerrymandering. To this purpose, we consider the subdivision of a $21\times21$ lattice $S$ into $n=5$ districts, with population $P_S\approx4700$  and voter distributions as in Table \ref{tab0}. The population and vote thresholds are fixed to be $t = 0.05 \times P_S/n$ and $w = 5$ in all cases.

Due to the inherent randomness built into the population distribution we can use the districts output by the redistricting algorithms with spatial restrictions, developed in Section \ref{sec4} \& \ref{sec4b}, to study aspects of gerrymandering via a Monte Carlo approach. Specifically, we generate a large sets of gerrymandered territory for the case of five districts and undertake a statistical analysis to quantify the advantage which gerrymandering brings to the proponent party and other characteristics of gerrymandered territories.
 
Whilst Monte Carlo methods have been previously employed for detection of gerrymandering, for instance \citep{Herschlag,Herschlag2}, and also towards constructing fair districts, e.g.~\citep{Fifield}, to our knowledge this is the first study to apply a Monte Carlo approach to look to quantify general features of gerrymandering.

\subsection{Quantitative Impact of Different Gerrymandering Strategies.}
\label{3.2}

The primary aim of the gerrymanderer is to maximize the predicted number of districts won $\#_{\rm win }$ during an election. The secondary aims are to construct proponent districts with $N_D\gg0$ and to arrange for opponent districts with $N_D\approx 0$, such that the districts constructed are secure from potential swings in the voting preferences against them, and can capitalize on swings in their favour to pick up additional districts.

To assess the impact of gerrymandering we compare predicted election results for each of the gerrymandering  algorithms developed above to the non-partisan symmetric districting model outlined in Section \ref{secbbb} and illustrated in Figure \ref{Fig7}. The predicted election results of the symmetric districting model are given in Table \ref{tab2}. Taking first the SRFH Packing algorithm, and following a similar methodology to Section \ref{secbbb}, we generate 30 population distributions on $S$ and calculate the average net vote per district $N_{D_k}$ and the average number of district wins for the proponent ($\#_{\rm win }$):

\begin{table}[H]
\begin{center}
\def\str{\vrule height11pt width0pt depth7pt}
\begin{tabular}{| c | c | c | c | c | c  | c | c | c | c | c | c | c | c | c  | c | c |  c | c | c | }
    \hline\str
Model  
 &  $N_{D1}$ & $N_{D2}$ & $N_{D3}$ & $N_{D4}$ & $N_{D5}$ & $\#_{\rm win }$   \\  
\hline
\#1      
 & 13.8      &  5.9      &  3.5    & -6.3       &  -16.6           & 3           \\ 
\#2              
 & 15.8       &   7.2        &  5.7  &  -14.8       & -14.2         & 3        \\ 
\#3         
& 12.6      &  13.2        & 9.7  & -2.4     & -31.6          &3.2           \\ 
\#4             
& 24.9      &  26.0      & 5.9    & -0.2       & -45.7         & 2.8       
    \\ 
\hline
\end{tabular}
\vspace{-3mm}
\caption{Average district vote after SRFH Packing of Section \ref{srfh}. \label{r1}}
\end{center}
\end{table}
\vspace{-3mm}
 SRFH-packing is most effective when oppositely extreme voters are situated in close proximity to one another. This approach handles the voter distribution of model \#4 notably less well (as can be seen from the fact that $\#_{\rm win }<3$) and although favouring the proponent party, the opponent takes the majority of districts in 30\% of trials. 

In the case of gerrymandering via Saturation packing we find that, although blunt, it can be extremely effective in securing districts for territories with a balanced vote:
\begin{table}[H]
\begin{center}
\def\str{\vrule height11pt width0pt depth7pt}
\begin{tabular}{| c | c | c | c | c | c  | c | c | c | c | c | c | c | c | c  | c | c |  c | c | c | }
    \hline\str
Model  
 &  $N_{D1}$ & $N_{D2}$ & $N_{D3}$ & $N_{D4}$ & $N_{D5}$ & $\#_{\rm win }$   \\  
\hline
\#1      
 & -187      &  136      &  32  &   11      &  8          & 4         \\ 
\#2              
 &  -338    & 219   & 45  &  25    &   48       &  4       \\ 
\#3         
& -438    & 65       & 207   & 63      & 102           & 4           \\ 
\#4             
& -222    &  11    &  128   &  15    &  68           &   4      \\ 
\hline
\end{tabular}
\vspace{-3mm}
\caption{Average district vote after Saturation Packing  of Section \ref{sat}. \label{r2}}
\end{center}
\end{table}
\vspace{-3mm}
Finally, we consider applications of the Genetic Gerrymandering algorithm, after each run we label the districts in order of net vote (prior to calculating the district averages): 
\begin{table}[H]
\begin{center}
\def\str{\vrule height11pt width0pt depth7pt}
\begin{tabular}{| c | c | c | c | c | c  | c | c | c | c | c | c | c | c | c  | c | c |  c | c | c | }
    \hline\str
Model  
 &  $N_{D1}$ & $N_{D2}$ & $N_{D3}$ & $N_{D4}$ & $N_{D5}$ & $\#_{\rm win }$   \\  
\hline
\#1      
 &  36.9       &  12.3        &  0.8   &  -12.1    &   -37.5       & 2.7           \\ 
\#2              
 &   117      &   57        &   16    &   -54     &   -137          & 2.7     \\ 
\#3         
&  232      &  116       &  -4    &  113     &  232           & 2.5     \\ 
\#4             
&  130      &  53       & 5    &-59    &  -119     & 2.5          \\ 
\hline
\end{tabular}
\vspace{-3mm}
\caption{Average district vote after Genetic Gerrymandering algorithm of Section \ref{sec4b}. \label{r3}}
\end{center}
\end{table}
\vspace{-3mm}
 Since in each case above, the territorial vote is close to balanced, $N_S\approx0$, in a fair election the expected average number of proponent districts wins is $\#_{\rm win }\approx2.5$. This is seen to be the case in Table \ref{tab2} for symmetric districting with symmetric sources. However, as anticipated, for the gerrymandered districts generated by our algorithms one observes clear skews in favour of the proponent. Specifically, SFRH Packing (Table \ref{r1}) leads to the proponent always winning 60\% of  districts and Saturation Packing (Table \ref{r2}) results in an impressive 80\% of  district wins  for the proponent . Genetic Gerrymandering (Table \ref{r3}) is less effective but still can provide a proponent advantage for voter models with only two source points.

\subsection{Compactness Tests}
\label{3.3}

One of the most identifiable traits of gerrymandering is that it typically leads to highly non-convex or non-compact districts.  Motivated by this common trait, researchers \citep{Roeck,Schwartzberg,Oxtoby,Young,Niemi,Polsby,Chambers,Hodge} have proposed geometric measures that can potentially be used to identify gerrymandering. It should be noted that simple geometric tests, while intuitive, do have certain drawbacks. For instance, they often do not account for variations in district boundaries due to natural or legal borders (e.g.~coastlines or national borders) and do not incorporate information on population distributions or demographics. Regardless, it is interesting to calculate these geometric measures for the gerrymandered districts arising in our work, in order to ascertain which index values likely indicate aggressive gerrymandering.

We shall examine the most intuitive dimensionless gerrymandering measure: the {\em isoperimetric quotient} index \citep{Polsby}, defined as
\begin{equation}
I_D:=\frac{4\pi A}{P^2}~,
\end{equation}
 for a district $D$ of area $A$ and perimeter $P$. The index is normalized to the circle, and deviations of a district from a circle give $I_D<1$. In Table \ref{tab8} we give the average $I_D$ for each district produced via  Genetic Gerrymandering (labelling districts in ascending $N_D$ after each run) as well as the mean $I_D$  for all districts averaged over 30 runs.

\begin{table}[H]
\begin{center}
\def\str{\vrule height11pt width0pt depth7pt}
\begin{tabular}{| c | c | c | c | c | c  | c | c | c | c | c | c | c | c | c  | c | c |  c | c | }
    \hline\str
Model   
 &  $I_{D1}$ & $I_{D2}$ & $I_{D3}$ & $I_{D4}$ & $I_{D5}$ & $I_{\rm average}$  \\
\hline
  \#1   
  & 0.5 &0.46    & 0.35       & 0.40   &  0.42    & 0.43            \\ 
  \#2       
  & 0.41 &  0.37       &   0.37       &  0.40    &  0.48   & 0.41           \\ 
  \#3       
  & 0.45 & 0.39      &   0.37        & 0.39 & 0.5  &  0.42             \\ 
  \#4         
  & 0.37 &  0.40     &  0.41       &  0.40  & 0.46 &  0.41                \\ 
\hline
\end{tabular}
\caption{Isoperimetric quotient of districts produced by genetic gerrymandering. \label{tab8}}
\vspace{-3mm}
\end{center}
\end{table}

In order to demonstrate that the districts generated by the Genetic Gerrymandering algorithm could potentially be detected and constrained by isometric quotient tests, we compare the $I_D$ values of Table \ref{tab8} to the isometric quotients calculated for the idealized non-partisan symmetric districts of Section \ref{secbbb}, these are given in Table \ref{tab10}.
\begin{table}[H]
\begin{center}
\def\str{\vrule height11pt width0pt depth7pt}
\begin{tabular}{| c | c | c | c | c | c  | c | c | c | c | c | c | c | c | c  | c | c |  c | c | }
    \hline\str
Model   
 &  $I_{D1}$ & $I_{D2}$ & $I_{D3}$ & $I_{D4}$ & $I_{D5}$ & $I_{\rm average}$  \\  
\hline
  \#1 - \#4  
  & 0.40 & 0.71       & 0.71        & 0.71    & 0.71       & 0.65               \\ 
\hline
\end{tabular}
\vspace{-2mm}
\caption{Isoperimetric quotient indices for  idealized non-partisan symmetric districts \label{tab10}}
\end{center}
\end{table}
\vspace{-10mm}
Since the idealized non-partisan symmetric districts follow simple geometric construction rules, without reference to the voter distribution, the same $I_D$ value is found for models  \#1 - \#4 (up to negligible fluctuations in the population distribution).

From the above analysis we find that our gerrymandering method always leads to an average $I_D$ which is much lower (around $33\%$) than the symmetric `fair' districts. Moreover, some of the gerrymandered districts have very low $I_D$ values, for instance the average value for District 3 of model 1 was $I_{D3}^{(\#1)}\approx0.35$, which is less than half the isoperimetric quotient of the typical symmetric district with $I_D\approx0.71$. We conclude that gerrymandered districts constructed via our Genetic Gerrymandering algorithm typically lead to highly non-compact districts for realistic distributions of voters. 

The implications of our results are that if redistricting were restricted by compactness constraints, then the extent to which one can impact an election would be significantly curtailed, and many aggressive gerrymandering methods (in particular our own) would likely prove to be ineffective. Thus this provides further support for the need for democratic governments to place some form of compactness requirement on redistricting schemes, and complements similar arguments and findings presented in e.g.~\citep{Schwartzberg,Polsby,Altman,Hodge,Humphreys,CC,Bowen,Chen}


		\section{Summary and Conclusions} 	
\label{sec6}

	The assignment of power to the electorate is the fundamental strength of representative democracies. Political gerrymandering, if left unchecked, threatens to undermine democratic systems. Since  gerrymandering issues can be formulated as well-stated mathematical problems, one can develop tools and tests to identify and inhibit such practices. 

We have presented a number of tools for studying gerrymandering and used these to examine central questions to gerrymandering. A lattice model for population distributions was developed and we proposed several algorithms to partition lattice territories  into gerrymandered, equal-population, connected (or mostly connected) districts. We used our lattice model to argue that the method of \cite{FH} is unrealistic since it generally leads to disconnected  districts. However, the techniques of packing and cracking remain central tools for gerrymandering, and we carried these ideas in constructing our own algorithmic approaches. To this end we developed three novel gerrymandering algorithms, SRFH Packing, Saturation Packing and Genetic Gerrymandering, and example executions of these strategies are shown in Figures \ref{Fig6b}, \ref{Fig6c}, \& \ref{Fig6d}. 

The probabilistic population fluctuations inherent to our voter models, allowed us to employ Monte Carlo methods to study the effectiveness of our gerrymandering algorithms.  We found that Saturation Packing, the blunt strategy of packing opposition voters into a small number of districts,  provided the most effective manner to secure district for a balanced vote. Moreover, our Saturation Packing algorithm led to a less fractured final district than SRFH Packing, with around 25\% fewer components. Genetic Gerrymandering was highlighted as our most complete algorithm, since it outputs districts which are connected with balanced populations as typically required to be legally valid, however the price for guaranteed connectedness in all districts is a significant reduction in partisan advantage for the gerrymander. Indeed, the Genetic Gerrymandering algorithm did not always assure electoral victory for a balanced vote, but it can tilt the territory in favour of the proponent for simple voter distributions. These weaker results are likely due to the genetic algorithm converging prematurely to local optima, and in future work we will confront this issue to construct more aggressive genetic gerrymandering algorithms.
 
	There are several other extensions of our algorithms which would be interesting to explore, such as incorporating third-party candidates and nonvoting electors. Furthermore, our current algorithms assume complete knowledge about the population distribution and voter preferences, but in reality, a gerrymanderer has imprecise information about these distributions, which leads to uncertainty in the voter preference distributions. Voter uncertainty can be implemented by allowing each $v_{i,j}$ to have some probabilistic fluctuations. One might also implement `voter shocks' to model swings in the popular vote by $\mathcal{O}(10\%)$ after redistricting is complete, as discussed in \cite{FH}. Both of these extensions directly affect the required vote threshold $w$, since if $w$ during redistricting is not sufficiently large, the proponent party may lose the election. Additionally, it would be interesting to extend the work initiated here to a more systematic study of different population distributions and spreads of partisan bias, in order to identify voter distributions which may be more vulnerable to gerrymandering. 
		
	We used the outputs of our gerrymandering algorithms to empirically demonstrate that gerrymandering provides a clear advantage for the gerrymanderer's party, and that the partisan connected districts output by our algorithms typically fail isoperimetric quotient tests.  Notably, while connectivity is legally mandated, compactness is not. Thus our work adds further support for implementing some legally mandated compactness requirement for redistricting to inhibit political gerrymandering.


\section*{\bf References}

\begingroup
\renewcommand{\section}[2]{}
	
	\endgroup
	
\end{document}